\documentclass[twocolumn,amsmath,amssymb,superscriptaddress,aps,pra]{revtex4}
\usepackage{amsmath,amsthm,amssymb}
\usepackage{latexsym}
\usepackage{amscd,graphicx}
\usepackage[titletoc]{appendix}
\usepackage{epsfig}
\usepackage{epstopdf}
\usepackage[normalem]{ulem}
\usepackage[dvipsnames]{xcolor}
\usepackage[colorlinks=true,linkcolor=red,citecolor=blue]{hyperref}
\usepackage{csquotes}

\begin{document}
	\title{Mechanical dynamics around higher-order exceptional point in magno-optomechanics}
	
	\author{Wen-Di He}
	\affiliation{Department of Physics, Wenzhou University, Zhejiang 325035, China}
	
	\author{Xiao-Hong Fan}
	\affiliation{Department of Physics, Wenzhou University, Zhejiang 325035, China}
	
	\author{Ming-Yue Liu}
    \affiliation{Department of Physics, Wenzhou University, Zhejiang 325035, China}
    
    \author{Guo-Qiang Zhang}
    \altaffiliation{zhangguoqiang@csrc.ac.cn}
    \affiliation{School of Physics, Hangzhou Normal University, Hangzhou 311121, China}
    
    \author{Hai-Chao Li}
    \altaffiliation{hcl2007@foxmail.com}
    \affiliation{College of Physics and Electronic Science, Hubei Normal University, Huangshi 435002, China}

	\author{Wei Xiong}
	\altaffiliation{xiongweiphys@wzu.edu.cn}
	\affiliation{Department of Physics, Wenzhou University, Zhejiang 325035, China}
		
\date{\today }
	
\begin{abstract}
We theoretically study diverse exceptional points (EPs) in an experimentally feasible magno-optomechanics consisting of an optomechanical subsystem coupled to a magnomechanical subsystem via physically direct contact. By adiabatically eliminating both the cavity and the Kittel mode, dissipative and parity-time symmetric exceptional points can be observed. When only the cavity mode is eliminated, a second (third)-order pseudo-Hermitian EP emerges for nondegenerate (degenerate) mechanical modes. The distinct dynamical behavior of two mechanical modes around these EPs are further studied.  Our proposal provides a promising way to engineer diverse EPs and quantify non-Hermitian phase transition with exceptional dynamical behavior in magno-optomechanics. 
\end{abstract}
	
	
\maketitle
	
\section{introduction}

Cavity optomechanics, which explores the interaction between photons in an optical cavity and phonons in a mechanical mode through radiation pressure, has been a vibrant field of theoretical and experimental research for over two decades~\cite{aspelmeyer2014cavity}. This field has enabled the investigation of various phenomena, including high-precision sensing~\cite{mason2019continuous}, ground-state cooling~\cite{PhysRevLett.110.153606,teufel2011sideband,groblacher2009demonstration,schliesser2008resolved}, quantum squeezing~\cite{PhysRevLett.114.093602,safavi2013squeezed,purdy2013strong}, optomechanically induced transparency~\cite{Optomechanically2010,safavi2011electromagnetically,PhysRevLett.111.133601,PhysRevLett.130.093603}, frequency conversion~\cite{PhysRevLett.108.153604,PhysRevLett.117.123902}, macroscopic entanglement~\cite{vitali2007optomechanical,PhysRevLett.110.233602,PhysRevLett.110.253601,PhysRevA.109.043512,PhysRevB.108.024105,liu2024tunable}, multistability~\cite{PhysRevLett.112.076402,PhysRevA.93.023844}, quantum criticality~\cite{PhysRevB.103.174106,Chen:21,lu_quantum-criticality-induced_2013,tian2023critical,PhysRevA.107.033516}, and quantum phase transitions~\cite{PhysRevLett.132.053601,PhysRevLett.120.063605,PhysRevLett.123.053601}. 

In parallel, magnons (i.e., collective spin excitations in ferromagnetic materials), with low dissipation~\cite{huebl2013high,tabuchi2014hybridizing}, excellent tunability~\cite{zhang2017observation,wang2016magnon}, and intrinsic Kerr effects~\cite{wang2018bistability,xiong2022strong,fan2024nonreciprocal,xiong2023highlytunable,zhang2019theory,PhysRevA.108.033704}, have emerged as key elements in quantum information science and condensed matter physics~\cite{doi:10.1126/sciadv.1501286,ZHANG2023100044,Zuo_2024,rameshti2022cavity,yuan2022quantum}. With the revelation of the magnetostrictive effect, magnomechanics is established, allowing to investigate macroscopic quantum entanglement~\cite{li2018magnon,li2019entangling,PhysRevA.109.043512,PhysRevB.108.024105}, nonclassical state~\cite{li2019squeezed,PhysRevA.109.013704,PhysRevA.107.063714}, frequency comb~\cite{PhysRevLett.131.243601,PhysRevA.107.053708}, and quantum networks~\cite{PRXQuantum.2.040344}.

To harness the advantages of both optomechanics and magnomechanics, hybrid magno-optomechanical systems have been recently proposed and experimentally demonstrated~\cite{PhysRevLett.129.243601}. These systems integrate optomechanical and magnomechanical components through the direct coupling of their mechanical modes, enabling microwave-to-optical conversion and mechanical interference between optically and microwave-driven motions~\cite{PhysRevLett.129.243601}. Quantum systems inevitably interact with their surrounding environment, transforming Hermitian systems into non-Hermitian ones. This provides an opportunity to study phase transition associated with the $n$th-order exceptional point (EPn), which have been extensively investigated in optomechanics~\cite{xu2016topological,jing2017high,zhang2018phonon,PhysRevA.104.063508,doi:10.1126/sciadv.abp8892,PhysRevA.106.033518,PhysRevLett.127.273603} and magnomechanics~\cite{lu2021exceptional,wang2023exceptional,wang2019magnon,huai2019enhanced}.
{Around EP2, lots of fascinating phenomena like unidirectional invisibility~\cite{chang2014parity,peng2014parity,lin2011unidirectional}, single-mode lasing~\cite{feng2014single,hodaei2014parity}, sensitivity enhancement~\cite{chen2017exceptional,hokmabadi2019non}, energy harvesting~\cite{fernandez2021enhanced}, protecting the classification of exceptional nodal topologies~\cite{staalhammar2021classification}, electromagnetically induced transparency~\cite{guo2009observation,wang2020electromagnetically,wang2019mechanical,lu2021exceptional}, and quantum squeezing~\cite{pevrina2019nonclassical,mukherjee2019enhancement,luo2022quantum} can be studied. Compared to EP2,  higher-order EPs, such as EP3 (third-order EP), can exhibit greater advantages in spontaneous emission enhancement~\cite{lin2016enhanced}, sensitive detection~\cite{hodaei2017enhanced,zeng2021ultra,wang2021enhanced,zeng2019enhanced}, topological characteristics~\cite{ding2016emergence,delplace2021symmetry}.}
However, EPs in magno-optomechanics remain unexplored, even though these systems offer much greater flexibility in parameter tuning. 
 
In this work, we propose to realize diverse EPs and study mechanical dynamics around these within magno-optomechanics. Firstly, we adiabatically eliminate the cavity and Kittel modes to obtain an effective mechanical subsystem. When two mechanical modes are dissipative, a dissipative EP2 can be predicted, where two complex eigenvalues coalesce. Around this point, The mechanical displacements oscillate dissipatively 
and decrease exponentially. When two mechanical modes are gain-loss balanced, a parity-time (PT) symmetric EP2 can be observed. Around PT symmetric EP2, two mechanical modes exhibit distinct dynamical behavior. Secondly, we only adiabatically eliminate the cavity mode for obtaining an effective three-mode magnomechanics. Under the pseudo-Hermitian condition, it can only host a pseudo-Hermitian EP2 when two mechanical modes are nondegenerate. But when two mechanical modes become degenerate, a pseudo-Hermitian EP3 emerges. The dynamical behavior of two mechanical modes around the pseudo-Hermitian EP2 and EP3 are further given. The results indicate that magno-optomechanical systems are promising platforms to realize higher-order EPs and the dyanmical behavior around EPs can be used to quantify non-Hermitian phase transition.


\section{Model and Hamiltonian}

\begin{figure}
	\includegraphics[scale=0.7]{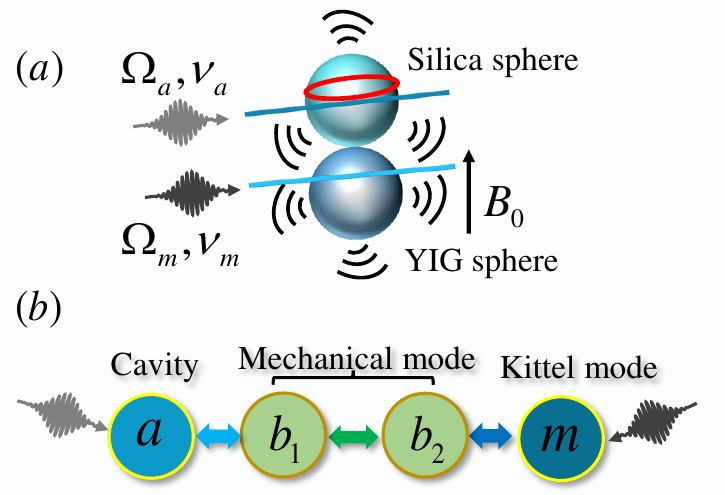}
	\caption{ (a) Schematic of the hybrid magno-optomechanical system consisting of a driven silica microsphere (optomechanical subsystem) coupled to a driven YIG microsphere (magnomechanical subsystem) located in a static magnetic field $B_0$ via directly physical contact. (b) Coupling configuration. The cavity mode is coupled to the mechanical mode $b_1$ via the optomechanical radiation pressure. The Kittel mode of the YIG sphere is coupled to the mechanical mode $b_2$ via the magnetostrictive force. Two mechanical modes are coupled via direct contact.}\label{fig1}
\end{figure}

We consider an experimentally feasible hybrid system~\cite{PhysRevLett.129.243601}, comprising a silica microsphere serving as the optomechanical cavity with angular frequency $\omega_a$ and a Kittel mode of a YIG microsphere functioning as the magnomechanical cavity with angular frequency $\omega_m$, where the {red}{optomechanical} cavity (Kittel) mode is driven by the external fields with angular frequency $\nu_a$ ($\nu_m$) and Rabi frequency $\Omega_a$ ($\Omega_m$), as illustrated in Fig.~\ref{fig1}(a). The coupling configuration is shown in Fig.~\ref{fig1}(b). The mechanical mode $b_1$ with angular frequency $\omega_1$ is radiately coupled to the cavity with the coupling strenght $g_a$, and the mechanical mode $b_2$ with angular frequency $\omega_2$ is magnetostrictively coupled to the Kittel mode with the coupling strength $g_m$. Two subsystems (i.e., optomechanics and magnomechanics) are coupled together via the straightway physical contact, allowing the coupling between two mechanical modes with coupling strength $J$. Under the strong driving fields, the proposed magno-optomechanics can be linearized following the standared approach. The linearized Hamiltonian reads ($\hbar = 1$)
\begin{align}\label{eq1}
H =&(\tilde{\Delta}_a-i\kappa_a) a^\dag a+G_a (a^\dag b_1+a b_1^\dag)\notag\\
&+(\tilde{\Delta}_m-i\kappa_m) m^\dag m+G_m (m^\dag b_2+m b_2^\dag),\\
&+\omega_1 b_1^\dag b_1+\omega_2 b_2^\dag b_2+J(b_1^\dag b_2+b_1 b_2^\dag),\notag
\end{align}
where $\tilde{\Delta}_{a(m)}=\omega_{a(m)}-\nu_{a(m)}+2g_{a(m)}{\rm Re}[b_{1(2),s}]$ is the effective frequency detuning shifted by the displacement of the mechanical mode $b_1$ ($b_2$), $G_{a(m)}=g_{a(m)} a_s (m_s)$ is the enhanced linearized optomechanical (magnetostrictive) coupling by the factor of $a_s$ ($m_s$), $\kappa_{a(m)}$ is the decay rate of the cavity (Kittel) mod, $a~(a^\dag),~m~(m^\dag),~b_1~(b_1^\dag)$, and $b_2~(b_2^\dag)$ are the annihilation (creation) operators of the cavity, the Kittel mode, the mechanical mode $b_1$, and the mechanical mode $b_2$, respectively. Note that tunable parameters $G_a$ and $G_m$ are assumed to be real hereafter, which can be realized by tuning the reference phases of two driving fields. Since the decay rates of the mechanical modes are much smaller compared to that of the cavity or Kittel mode, two mechanical modes are assumed to be neutral for convenience.

\section{Exceptional dynamics around EP2}

The dynamics of the system governed by Eq.~(\ref{eq1}) can be given by the Heisenberg equation of motion, 
\begin{align}\label{eq4}
	\dot{a}=&-(\kappa_a+i\tilde{\Delta}_a)a-iG_a b_1,\notag\\
	\dot{m}=&-(\kappa_m+i\tilde{\Delta}_m)m-iG_m b_2,\notag\\
	\dot{b}_1=&-i\omega_1 b_1-iG_a a-iJ b_2,\\
	\dot{b}_2=&-i\omega_2 b_2-iG_m m-iJ b_1.\notag
\end{align}
When $|\kappa _a|\gg G_a$ and $|\kappa _m| \gg G_m$, one can adiabatically eliminate both the cavity and Kittel modes. This can be realized via replacing the operators $a$ and $m$ in the last two equations in Eq.~(\ref{eq4}) by the expectation values $\langle a\rangle=-i(G_a/\kappa_a) b_1$ and $\langle m\rangle=-i(G_m/\kappa_m) b_2$, respectively, where $|\kappa _a|\gg\tilde{\Delta}_a$ and $|\kappa _m|\gg\tilde{\Delta}_m$ are further taken. {To satisfy these conditions, the accessible parameters $\omega_b/2\pi=10$ MHz, $\tilde{\Delta}_a=\tilde{\Delta}_m=0.32{\omega _b}$, 
	$\left| {{\kappa _a}} \right| = \left| {{\kappa _m}} \right|=\kappa ={\omega _b}$, ${G_a} = {G_m}=G = 0.32{\omega _b}$ are taken~\cite{li2018magnon,li2019entangling,PhysRevLett.129.243601,li2019squeezed}.} Hence, the dynamics in Eq.~(\ref{eq4}) reduces to
\begin{align}\label{eq5}
		\dot{b}_1=&-(\Gamma_1+i\omega_1)b_1-iJ b_2,\notag\\
	\dot{b}_2=&-(\Gamma_2+i\omega_2)b_2-iJ b_1.
\end{align}
The corresponding effective Hamiltonain is
\begin{equation}\label{eq7}
	H_{m} = \left( 
	\begin{array}{*{20}{c}}
		\omega_1-i\Gamma_1 & J \\
		J & \omega_2-i\Gamma_2
	\end{array} 
	\right).
\end{equation}
where $\Gamma_1 ={G_a^2}/{\kappa _a}$ and $\Gamma _2={G_m^2}/{\kappa _m}$ are effective and tunable decay rates of two mechanical modes, respectively.

\begin{figure}
	\includegraphics[scale=0.185]{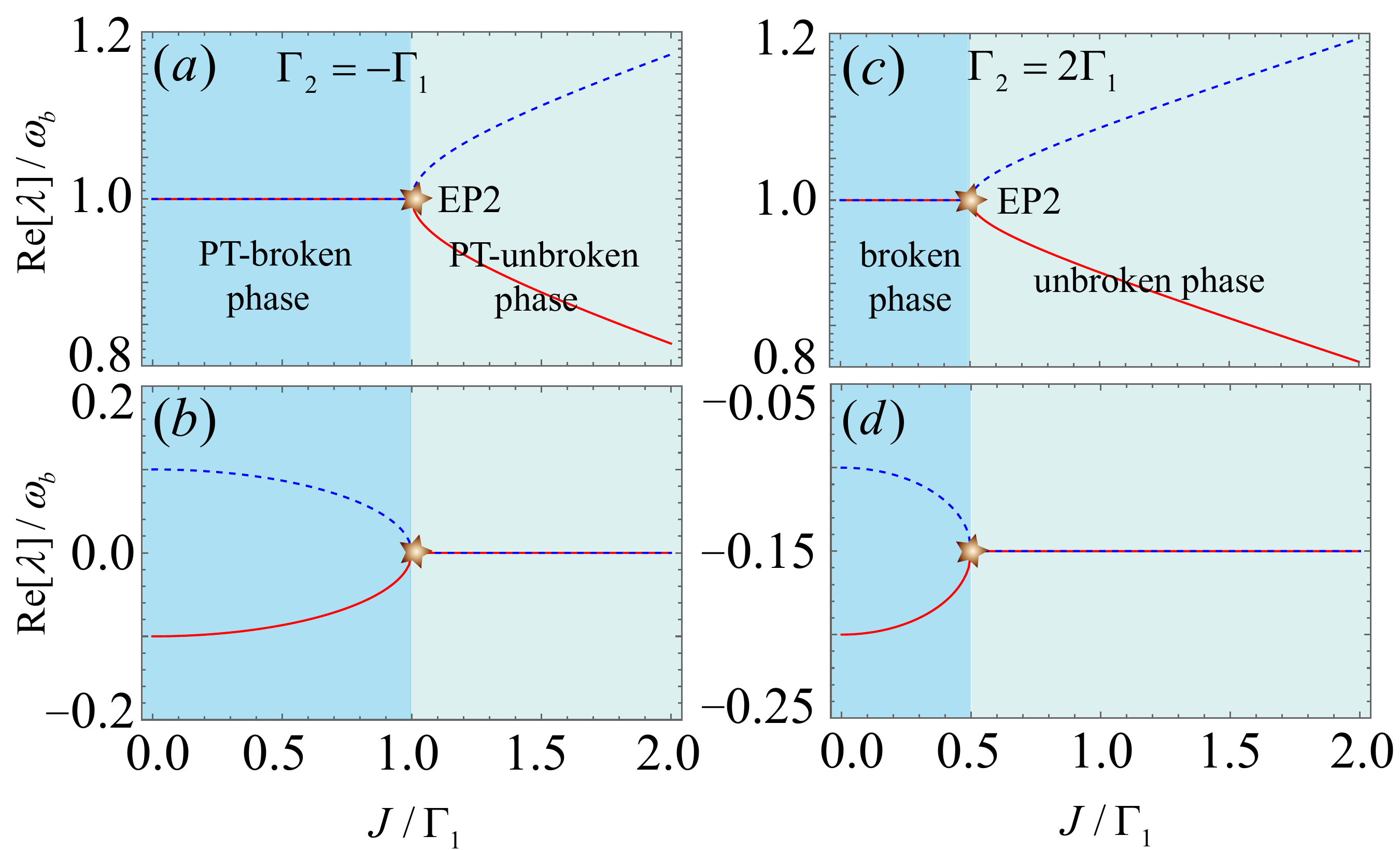}
	\caption{ Eigenvalues of the effective mechanical system governed by Eq.~(\ref{eq7}) vs the normalized coupling strength $J/\Gamma_1$. (a) and (b) PT-symmetric case ($\Gamma_2=-\Gamma_1$); (c) and (d) Purely dissipative case ($\Gamma_2=2\Gamma_1$). The parameters are chosen as $\omega_b/2\pi=10$ MHz and $\Gamma_1=0.1\omega_b$. }\label{fig2}
\end{figure}

The eigenvalues of the Hamiltonian $H_m$ can be given by 
\begin{align}\label{eq8}
\lambda_\pm=&\omega_b\pm\Omega-i\Gamma_+=\omega_\pm-i\Delta\omega_\pm,
\end{align}
where $\omega_b=\omega_1=\omega_2$,  $\Omega=\sqrt{J^2-\Gamma_-^2}$, and $\Gamma_\pm=(\Gamma_1\pm\Gamma_2)/2$. $\omega_\pm$ and $\Delta\omega_\pm$ respectively denote the eigenfrequencies and linewidths of two supermodes hybridized by the coupling between two mechanical modes. It is evident that the mechanical system exhibits PT symmetry when $\Gamma_1=-\Gamma_2$ ($\Gamma_+=0$), i.e., $[{PT},H_m]=0$, where $P$ is the parity operator and $T$ is the time operator. To achieve PT symmetry, one mechanical mode must be lossy while the other must be gainy. Without loss of generality, we assume the mechanical mode $b_1$ is lossy ($\Gamma_1>0$) and the mechanical mode $b_2$ is gainy ($\Gamma_2<0$). This can be realized by employing a lossy cavity and a gainy Kittel mode. By further imposing $\Omega^2=0$, i.e., $J=\Gamma_-$, a PT symmetric EP2 emerges [see stars in Figs.~\ref{fig2}(a) and \ref{fig2}(b)], where two eigenvalues coalesce, $\omega_+=\omega_-=\omega_b$ and $\Delta\omega_+=\Delta\omega_-=0$. This indicates that high-quality supermodes can be obtained.
When $\Omega^2<0$, i.e., $J<\Gamma_-$, the system enters the PT-broken phase [see blue regions in Figs.~\ref{fig2}(a) and \ref{fig2}(b)]. In this phase, two supermodes are resonant ($\omega_+=\omega_-=\omega_b$), but have opposite linewidths ($\Delta\omega_+=-\Delta\omega_-=0.1\omega_b$). This indicates that one supermode is lossy and the other is gainy. In the PT-unbroken phase~[see light green regions in Figs.~\ref{fig2}(a) and \ref{fig2}(b)], i.e., $\Omega^2>0$ or $J>\Gamma_-$, two supermodes with zero linewidth ($\Delta\omega_+=\Delta\omega_-=0$) are detuned ($\omega_+\neq\omega_-$). 
\begin{figure}
	\includegraphics[scale=0.4]{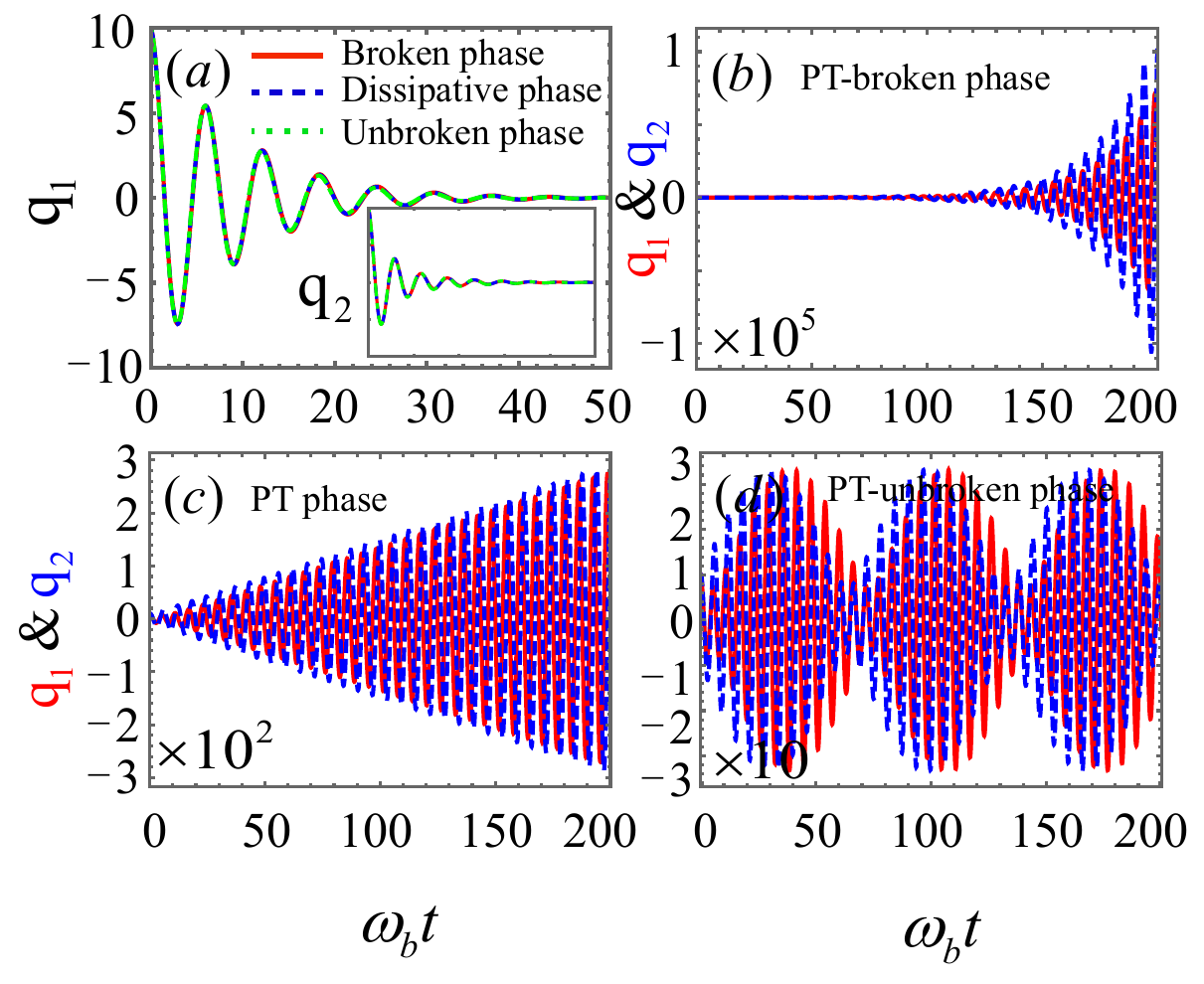}
	\caption{The dynamical behavior of the two mechanical modes near (a) dissipative EP2 and (b,c,d) PT symmetric EP2. Other parameters are the same as those in Fig.~\ref{fig2}.}\label{fig3}
\end{figure}
When $\Gamma_1>0$ and $\Gamma_2>0$, the system described by $H_m$ in Eq.~(\ref{eq7}) is purely dissipative. For $\Omega^2=0$ (or $J=|\Gamma_-|$), a dissipative EP2 emerges, demonstrated by Figs.~\ref{fig2}(c) and \ref{fig2}(d). For $\Omega^2<0$ (or $J<|\Gamma_-|$), the system enters the broken phase [see blue regions in Figs.~\ref{fig2}(c) and \ref{fig2}(d)], where two supermodes have the same eigenfrequency ($\omega_+=\omega_-=\omega_b$) but different linewidths ($\Delta\omega_+\neq\Delta\omega_-$). When $\Omega^2>0$ (i.e., $J>-\Gamma_-$), eigenfrequencies of two supermodes are detuned but linewidths are identical in the unbroken phase [see light green regions in Fig.~\ref{fig2}(c) and \ref{fig2}(d)].

To study the dynamical behavior of two mechanical modes around disspative and PT symmetric EP2s, we rewrite Eq.~(\ref{eq5}) as
\begin{align}\label{eq9}
	\dot{q}_1=&-\Gamma_1q_1+\omega_b p_1+J p_2,\notag\\
	\dot{p}_1=&-\Gamma_1p_1-\omega_bq_1-J q_2,\notag\\
	\dot{q}_2=&-\Gamma_2q_2+\omega_b p_2+J p_1,\\
	\dot{p}_2=&-\Gamma_2p_2-\omega_bq_2-J q_1,\notag
\end{align}
where $q_{1(2)} = (b_{1(2)}^\dag  + b_{1(2)})/\sqrt 2$ and $p_{1(2)} = i(b_{1(2)}^\dag  - b_{1(2)})/\sqrt 2$ are defined. With $p_1(t=0)=p_2(t=0)=0$, $q_1(t)$ and $q_2(t)$ can be given by
\begin{align}\label{eq10}
{q_1}(t) = &e^{-\Gamma_+t}\bigg\{ \bigg[\cos ({\Omega t})-\frac{2\Gamma_-}{\Omega}\sin ({\Omega t})\bigg]\cos(\omega_b t)q_1(0)\notag\\
 &-\frac{2J}{\Omega}\sin ({\Omega t})\sin (\omega_b t)q_2(0)\bigg\}, \\
{q_2}(t) = &e^{-\Gamma_+t}\bigg\{\bigg[\cos ({\Omega t})-\frac{2\Gamma_-}{\Omega}\sin ({\Omega t})\bigg]\cos(\omega_b t)q_2(0)\notag\\
&-\frac{2J}{\Omega}\sin ({\Omega t})\sin (\omega_b t)q_1(0)\bigg\}, \notag
\end{align}
which clearly reveals that $q_1$ and $q_2$ (see the inset) have the same dynamical behavior near dissipative EP2, as shown in Fig.~\ref{fig3}(a). Specifically, both $q_1$ and $q_2$ dissipatively oscillate and exponentially decrease.

When $\Gamma_1=-\Gamma_2$, i.e., $\Gamma_+=0$, the mechanical system is PT symmetry, leading the exponential factor in Eq.~(\ref{eq10}) to vanish. Therefore, the dynamical behavior of the two mechanical modes are distinct near PT symmetric EP2, as shown in Figs.~\ref{fig3}(b-d). In the PT-broken phase [see Fig.~\ref{fig3}(b)], two mechanical modes are initially stationary. As time goes, two mechanical modes begin to oscillate and increase exponentially. {This is because the term $\Omega$ in Eq.~(\ref{eq8}) is imaginary in the PT-broken phase, thus the function $\exp(-i\Omega t)$ becomes an exponetial function, that is, the cosine and sine functions related to $\Omega$ in Eq.~(\ref{eq10}) actually are hyperbolic time functions, leading to an exponentially oscillation. Physically, the imaginary $\Omega$ indicates that the system with the eigenvalue $\lambda_+$ is gainful. Thus, the energy is continuously injected from the environment into the system, leading to instability and causing exponential oscillation.}  Morever, $q_2$ increases faster than $q_1$.  In the PT phase [see Fig.~\ref{fig3}(c)], $q_1$ and $q_2$ initially oscillate with tiny but the equal amplitudes and increase linearly with the time. In the PT-unbroken phase [see Fig.~\ref{fig3}(d)], $q_1$ and $q_2$ oscillate periodically, forming beat frequency patterns. {This is due to the fact that the term $\Omega$ in Eq.~(\ref{eq8}) is real, so the function $\exp(-i\Omega t)$ is a cosine or sine function of $\Omega t$, leading to periodical oscillation, which can also be obtained from Eq.~(\ref{eq10}). Physically, the real part of the eigenvalue can only give rise to a phase during the time evolution.} These significant differences in the dynamical behaviors of the two mechanical modes can be used to probe the phase transition from the PT-broken phase to the PT-unbroken phase, or vice versa. 

\section{Exceptional dynamics around pseudo-Hermitian EP3 and EP2}

\begin{figure}
	\includegraphics[scale=0.3]{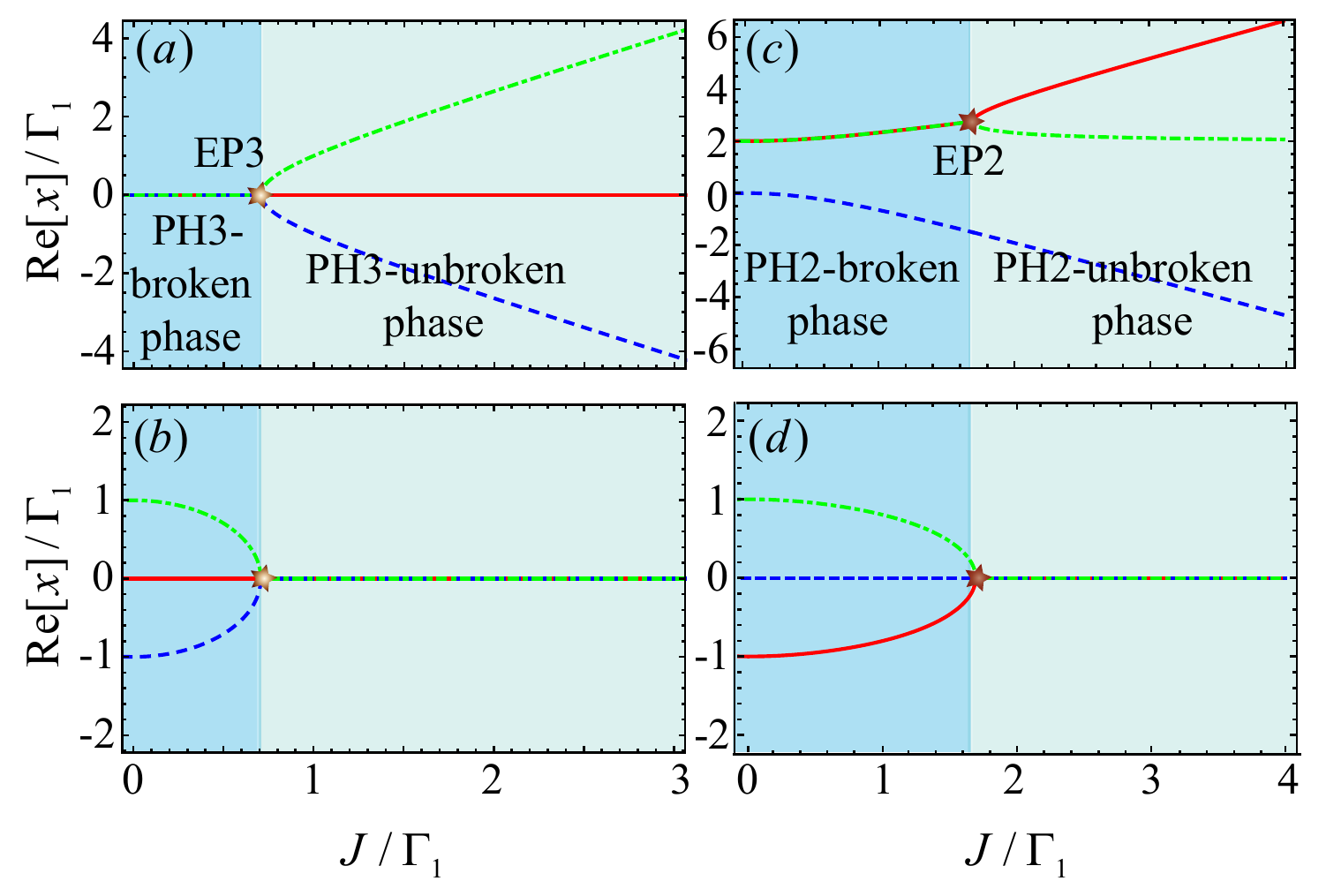}
	\caption{Eigenvalues of the Hamiltonian $H_{mb}$ vs the normalized coupling strength $J/\Gamma_1$. (a) and (b) Degenerate mechanical modes ($\delta=0$); (c) and (d) Nondegenerate mechanical modes ($\delta=2\Gamma_1$). PH3 (PH2) denote the pseudo-Hermitian EP3 (EP2). Other parameters are the same as those in Fig.~\ref{fig2}.}\label{fig4}
\end{figure}
When the decay rate of the cavity mode is much larger than the optomechanical coupling in Eq.~(\ref{eq4}), i.e., $\kappa_a \gg G_a$, the cavity mode can be adiabatically eliminated, then Eq.~(\ref{eq4}) becomes
\begin{align}\label{eq12}
	\dot{m}=&-(\kappa_m+i\tilde{\Delta}_m)m-iG_m b_2,\notag\\
	\dot{b}_1=&-(\Gamma_1+i\omega_1)b_1-iJ b_2,\\
	\dot{b}_2=&-i\omega_2 b_2-iJ b_1-i G_m m.\notag
\end{align}
The corresponding non-Hermitian Hamiltonian in the matrix form reads
\begin{equation}\label{eq14}
H_{mb} = \left( 
\begin{array}{*{20}{c}}
		\tilde{\Delta}_m-i\kappa_m & 0 & G_m\\
		0 & \omega_1 - i\Gamma _1 & J \\
		G_m & J & \omega_2
\end{array} 
\right).
\end{equation}
Obviously, the non-Hermitian Hamiltonian $H_{mb}$ has three eigenvalues. When these three eigenvalues are real or one of the three eigenvalues is real and the other two are a complex conjugate pair, the Hamiltonian $H_{mb}$ is pseudo-Hermitian. For a pseudo-Hermitian system, solutions of the characteristic equation $\left| H_{mb}- I\Lambda\right| = 0$ are the same as that of $\left| H_{mb}^*-I \Lambda \right| = 0$, with $\Lambda$ being the eigenvalue and $I$ an identity matrix. By comparing these two characteristic equations, the pseudo-Hermitian condition can be directly obtained as
\begin{align}\label{eq15}
	\Gamma _1 =-\kappa_m,~\tilde{\Delta}_m=\omega_1,~G_m=J.
\end{align}
The first equality indicates that the gain of the Kittel mode and the effective loss of the mechanical mode $b_1$ are required to be balanced. The second equality means that the magnomechanical system works on the red sideband of the mechanical mode $b_1$. The last equality reveals that uniform coupling strengths are needed. With the pseudo-Hermitian condition, the characteristic equation $\left| {H_{mb} -I \Lambda} \right| = 0$ can be specifically expressed as
\begin{align}\label{s16}
	x^3-2\delta x^2	-(2J^2-\delta^2-\Gamma_1^2)x+2J^2\delta=0,
\end{align}
where $x=\Lambda- \omega_2$ and $\delta=\omega_1-\omega_2$. The solutions of this equation are $x_1=\frac{1}{3}(2\delta-a/c+c)$, $x_2=\frac{1}{3}(2\delta+\frac{1+i\sqrt{3}}{2c}a-\frac{1-i\sqrt{3}}{2}c)$, and $x_3=\frac{1}{3}(2\delta+\frac{1-i\sqrt{3}}{2c}a-\frac{1+i\sqrt{3}}{2}c)$,
with $a=-6J^2-\delta^2+3\Gamma_1^2,~b=-2(9J^2+9\Gamma_1^2+\delta^2)\delta$, and $c=(b/2+\sqrt{a^3+b^2/4})^{1/3}$.

When two mechanical modes are degenerate, i.e., $\delta=0$, three eigenvalues reduce to
\begin{align}
	x_1=0,~~x_2=\sqrt{2J^2-\Gamma_1^2},~~x_3=-\sqrt{2J^2-\Gamma_1^2}.
\end{align}
One can easily find that these three values coalesce at a pseudo-Hermitian EP3, where $\sqrt{2}J=\Gamma_1$, demonstrated by  
Figs.~\ref{fig4}(a) and \ref{fig4}(b). When $\sqrt{2}J<\Gamma_1$, the system is in the pseudo-Hermitian-broken phase [see the blue regions in Figs.~\ref{fig4}(a) and \ref{fig4}(b)], where eigenfrequencies of three supermodes are still degenerate, but the linewidths bifurcate (one supermode has zero linewidth, and the other two have opposite linewidths). In the pseudo-Hermitian-unbroken phase [see the light green regions in Figs.~\ref{fig4}(a) and \ref{fig4}(b)], eigenfrequencies bifurcate into three values, but linewidths are identical.

\begin{figure}
	\includegraphics[scale=0.32]{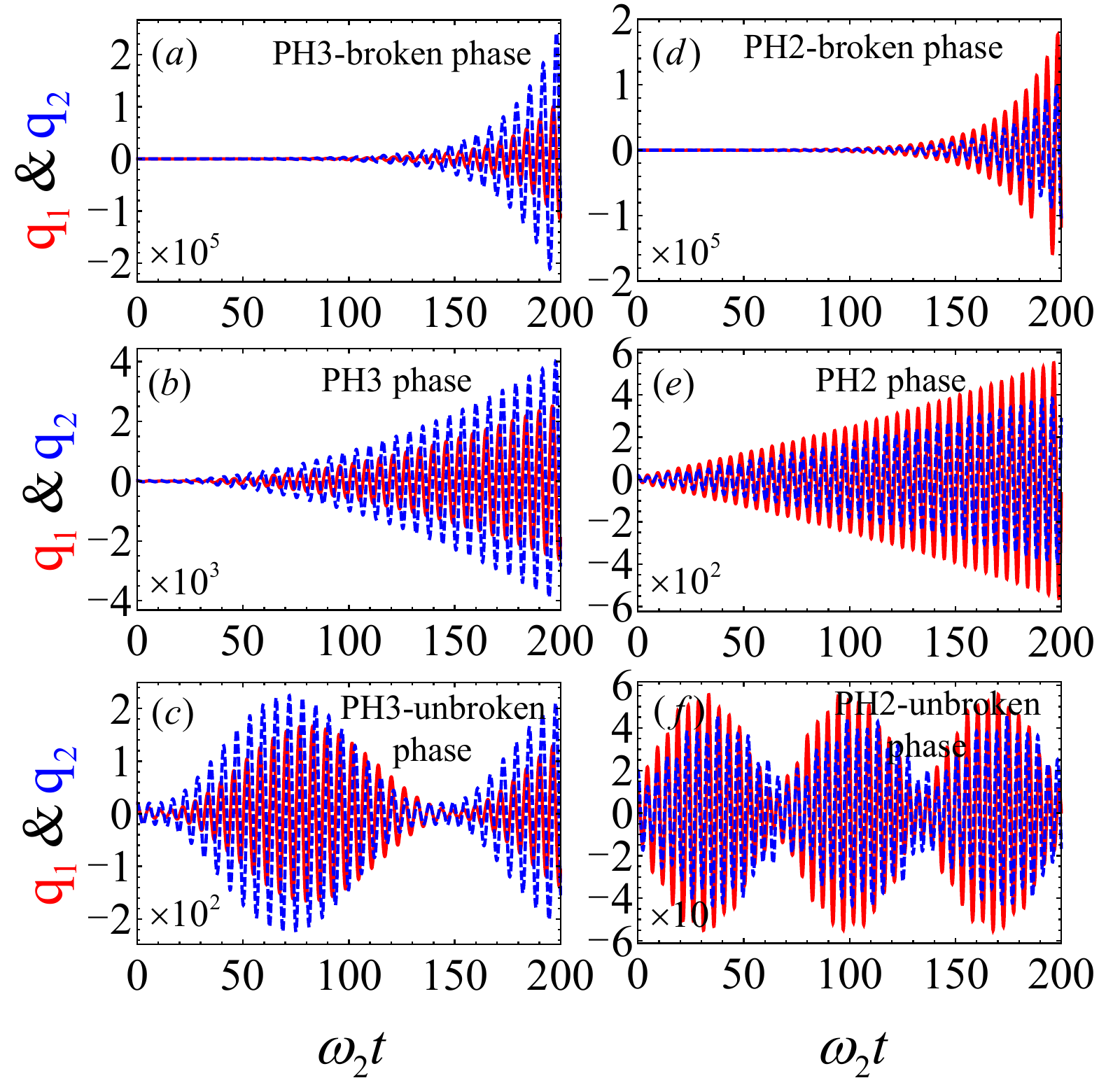}
	\caption{The dynamical behavior of the two mechanical modes in (a) PH3-broken phase ($\sqrt{2}J=0.9\Gamma_1$), (b) PH3 phase ($\sqrt{2}J=\Gamma_1$), (c) PH3-unbroken phase ($\sqrt{2}J=1.1\Gamma_1$), (d) PH2-broken phase ($J/\Gamma_1=1.52$), (e) PH2 phase ($J/\Gamma_1=1.69$), and (f) PH2-unbroken phase ($J/\Gamma_1=1.86$). Other parameters are the same as those in Fig.~\ref{fig2}. }\label{fig5}
\end{figure}

When two mechanical modes are nondegenerate, such as $\delta=2\Gamma_1$, we find that only two eigenvalues $x_1$ and $x_3$ coalesce at a pseudo-Hermitian EP2, where $J/\Gamma_1=1.69$ [see Figs.~\ref{fig4}(c) and \ref{fig4}(d)]. When $J/\Gamma_1<1.69$, the system enters the pseudo-Hermitian-broken phase, where the eigenfrequencies of the degenerate supermodes are identical, but the linewidths are opposite. When $J/\Gamma_1>1.69$, i.e., the pseudo-Hermitian-unbroken phase, the linewidths of the degenerate supermodes are identical, but the eigenfrequencies becomes bifurcation. Across the pseudo-Hermitian EP2 from the broken to unbroken phases, the eigenvalue $x_2$ is negative and increases with the coupling strength between two mechanical modes.

To further investigate the dynamical behavior of two mechanical modes around the pseudo-Hermitian EPs, we rewrite Eq.~(\ref{eq12}) as
\begin{align}\label{eq18}
	\dot{q_1}  =& \omega _1 p_1 + J p_2 - \Gamma _1 q_1,\notag\\
	\dot{q_2}  =& \omega _2 p_2 + J p_1 + G_m p_m,\notag\\
	\dot{q_m}  =& \tilde{\Delta}_m p_m - k_m q_m + G_m p_2,\notag\\
	\dot{p_1} = & - \omega _1 q_1 - J q_2 - \Gamma _1 p_1,\\
	\dot{p_2} =  &- \omega _2 q_2 - J q_1 - G_m q_m,\notag\\
	\dot{p_m}   =  &- \tilde{\Delta}_m q_m - k_m p_m - G_m q_2,\notag
\end{align}
where $q_m=(m^\dag+m)/\sqrt{2}$, $p_m=i(m^\dag-m)/\sqrt{2}$, and $q_{1(2)}$, $p_{1(2)}$ are the same as in Eq.~(\ref{eq9}). We numerically solve Eqs.~(\ref{eq18}) under the pseudo-Hermitian condition in Eq.~(\ref{eq15}) with initial conditions $q_1(0)=q_2(0)=2q_m(0)=20$ and $p_1(0)=p_2(0)=p_m(0)=0$ and show the result in \textcolor{red}{Fig.~\ref{fig5}}. When two mechanical modes are degenerate ($\delta=0$), only a pseudo-Hermitian EP3 can be predicted when $\sqrt{2}J=\Gamma_1$. At this point, $q_1$ and $q_2$ initially have weak oscillations. For a longer time, both increase exponentially but $q_2$ is much faster, confirmed by Fig.~\ref{fig5}(b). When $\sqrt{2}J<\Gamma_1$, the system is in the pseudo-Hermitian-broken phase [see Figs.~\ref{fig5}(a)], where $q_1$ and $q_2$ exhibit similar behavior to that at EP3, except for longer time needed to begin oscillation and the amplitudes are significant enhanced. In the pseudo-Hermitian-unbroken phase ($\sqrt{2}J>\Gamma_1$), $q_1$ and $q_2$ oscillate periodically, exhibiting beat frequency patterns, as demonstrated by Fig.~\ref{fig5}(c).

When two mechanical modes are nondegenerate ($\delta=2\Gamma_1$), only a pseudo-Hermitian EP2 can be observed for $J/\Gamma_1=1.69$, where $q_1$ and $q_2$ initially oscillate and increase exponentially [see Fig.~\ref{fig5}(e)], different from the dissipative and PT symmetric cases. In the pseudo-Hermitian-broken ($J/\Gamma_1<1.69$) and unbroken phases ($J/\Gamma_1>1.69$) [see Figs.~\ref{fig5}(d) and \ref{fig5}(f)], $q_1$ and $q_2$ have the similar dynamical behavior to the cases in the PT-broken and unbroken phases, respectively. However, the amplitude of $q_1$ is larger than that of $q_2$ near the pseudo-Hermitian EP2, which is fully opposite with the case near the PT symmetric EP2.

\begin{figure}
	\includegraphics[scale=0.18]{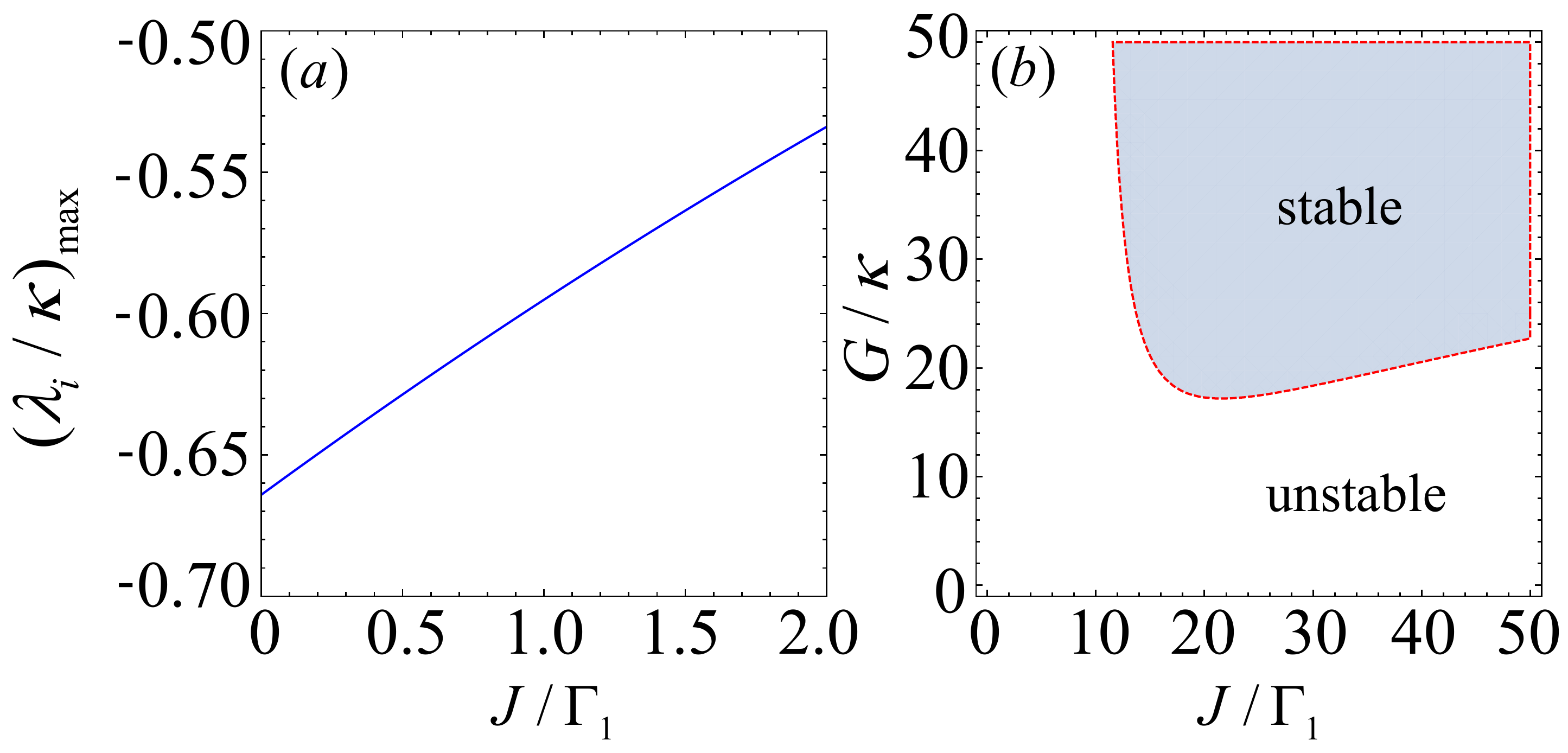}
	\caption{{(a) The maximal eigenvalue of the coefficient matrix of Eq.~(\ref{eq4}) vs the coupling strength $J$, where ${{\kappa _a}} =  {{\kappa _m}}=\kappa ={\omega _b}>0$ and ${G_a} = {G_m}=G = 0.32{\omega _b}$. (b) The stability diagram vs the coupling strengths $J$ and $G$, where $\kappa_a=-\kappa_m=\kappa={\omega _b}>0$. In (a) and (b), other parameters are $\omega_b/2\pi=10$ MHz, $\tilde{\Delta}_a=\tilde{\Delta}_m=0.32{\omega _b}$.} }\label{fig6}
\end{figure}

\section{Discussion and Conclusion}

{Before conclusion, we give a brief discussion on the stability of our proposed system via the Routh-Hurwitz criteri. When no gain is introduced, the system can be stable in a long time, as shown in Fig. 3(a). It can also be numerically demonstrated by plotting the maximal value of the coefficient matrix in Eq.~(\ref{eq4}) vs the coupling strength $J$ [see Fig.~\ref{fig6}(a)]. However, when the gain is introduced, the system can be stable when the enhanced optomechanical and magnetostrictive coupling strengths are comparable and both of them are much larger than the decay rate of the cavity. [see the light-blue region in Fig.~\ref{fig6}(b)]. This is because that the energy injected to the Kittel mode (i.e., gain) can be first transferred to the dissipative mechanical resonator via the optomechanical coupling and then to the disspative optomechanical system via the resonator-resonator coupling $J$, leading to the whole system stable. But when the optomechanical and magnetostrictive coupling strengths are mismatched or both of them are not very strong [see the white region in Fig.~\ref{fig6}(b)], the system is unstable. This is due to the fact that the energy injected into the Kittle mode can not be taken away by the dissipative resonator via the magnetostrictive coupling, leading to the whole system unstable. In fact, EPs in unstable or nearly unstable non-Hermitian systems have been investigated yet\cite{xu2015mechanical,xu2021optomechanical}. Further, some novel phenomena can be found within unstable non-Hermitian systems, such as uncovering inherent instability of large clusters~\cite{li2021non}, unstable population dynamics~\cite{martello2023coexistence}, phase transitions~\cite{rahmani2024exceptional}, nonexponential decay~\cite{jittoh2005nonexponential}, and noise-enhanced stability~\cite{mantegna1996noise}.}

In summary, we have theoretically study EP2 and EP3 in an experimentally feasible magno-optomechanics consisting of a magnomechanical system coupled to an optomechanical system via physically dicrect contact. By selectively eliminating the cavity mode or both the cavity and Kittel modes, diverse EPs including dissipative EP2, PT symmetric EP2, pseudo-Hermitian EP2 and EP3 can be predicted.  We further show that the mechanical dynamics around these EPs are significantly distinct. The result indicates that the proposed system is a promising platform to investigate diverse EPs and quantify non-Hermitian phase transition with mechanical dynamical behavior.

	
This work is supported by the Zhejiang Provincial Natural Science Foundation of China under Grant No. LY24A040004, the Natural Science Foundation of Hubei Province of China under Grant No. 2022CFB509, the National Natural Science Foundation of China (Grants No. 11904201 and No. 12205069).

\bibliography{ms2}
\end{document}